\documentclass[10pt,twocolumn,twoside]{IEEEtran}
\ifCLASSINFOpdf
\else
\fi
\usepackage{graphicx} 
\usepackage{epsfig} 
\usepackage{amsmath} 
\usepackage{amssymb}  
\usepackage{psfrag}
\usepackage{color}
\usepackage{cite}
\usepackage[T1]{fontenc}

\graphicspath{{Figures/}}

\newtheorem{thm}{Theorem} 
\newtheorem{prop}{Proposition} 
\newtheorem{lem}{Lemma}

\newtheorem{defi}{Definition}
\newtheorem{rem}{Remark}

\begin{document}
%
\title{
Minimum-Rank Dynamic Output Consensus Design for Heterogeneous Nonlinear Multi-Agent Systems   
}
%
%
%

\author{Dinh Hoa Nguyen, \IEEEmembership{Member, IEEE} 
\thanks{Dinh Hoa Nguyen is currently with Control System Laboratory, Department of Advanced Science and Technology, Toyota Technological Institute, 
        2-12-1 Hisakita, Tempaku-ku, Nagoya 468-8511, Japan. e-mail: dinhhoa\_nguyen@toyota-ti.ac.jp, hoadn.ac@gmail.com.} }

\newcounter{MYtempeqncnt}

\maketitle


\begin{abstract}
\textcolor{blue}{
In this paper, we propose a new and systematic design framework for output consensus in heterogeneous Multi-Input Multi-Output (MIMO) general nonlinear Multi-Agent Systems (MASs) subjected to directed communication topology. 
First, the input-output feedback linearization method is utilized assuming that the internal dynamics is Input-to-State Stable (ISS) to obtain linearized subsystems of agents. 
Consequently, we propose local dynamic controllers for agents such that the linearized subsystems have an identical closed-loop dynamics which has a single pole at the origin whereas other poles are on the open left half complex plane. This allows us to deal with distinct agents having arbitrarily vector relative degrees and to derive rank-$1$ cooperative control inputs for those homogeneous linearized dynamics which results in a minimum rank distributed dynamic consensus controller for the initial nonlinear MAS. Moreover, we prove that the coupling strength in the consensus protocol can be arbitrarily small but positive and hence our consensus design is non-conservative.  
Next, our design approach is further strengthened by tackling the problem of randomly switching communication topologies among agents where we relax the assumption on the balance of each switched graph and derive a distributed rank-$1$ dynamic consensus controller. 
Lastly, a numerical example is introduced to illustrate the effectiveness of our proposed framework.   
}
\end{abstract}




\section{Introduction}
%
%
%
%


 %

%
%

Cooperative control of multi-agent systems (MASs) has gained much attention recently since there are a lot of practical applications, e.g.,     
power grids, wireless sensor networks, transportation networks, systems biology, etc, can be formulated, analyzed and synthesized under the framework of MASs. 
One of the key features in MASs is the achievement of a global objective by performing local measurement and control at each agent and simultaneously collaborating among agents using that local information. 

Employing the principle of relatively exchanged local information, a very important and extensively studied subject in MASs is the consensus problem where agents' states or outputs come to a non-zero agreement. 
A huge collection of results can be found for consensus of MASs, ranging from single integrator dynamics of agents with fixed and time-varying communication topology 
\cite{Jadbabaie:2003,Olfati-Saber:2004,Olfati-Saber:2007,Ren:2007} to general linear agents \cite{F.Xiao:2007} and to nonlinear agents with disturbances, uncertainties, time delays, etc 
\cite{Chopra:2006,Su-Huang:2015}.    

For linear MASs, one way to develop a systematic consensus control design is to employ the LQR method, e.g. \cite{Cao:2010,Zhang:2011,Nguyen:2014,Nguyen:2015TAC1}.    
The paper \cite{Cao:2010} designed optimal consensus laws for network of integrators utilizing two LQR cost functions.  
LQR-based consensus designs for leader-follower MASs was presented in \cite{Zhang:2011} in which only local LQR problems were solved and no global LQR problem was considered.  
Next, in our previous research \cite{Nguyen:2014}, we introduced an LQR-based method to design a distributed consensus controller for general linear MASs but the obtained controller is only sub-optimal. 
Recently, we have proposed an approach in \cite{Nguyen:2015TAC1} to achieve a consensus design with a non-conservative coupling strength where an alternative MAS model namely edge dynamics was presented that helps transforming the consensus design into an equivalent stability synthesis which can be derived by LQR method. This advanced result will be employed subsequently in the current article.     

In nonlinear MASs, many efforts have recently been conducted for consensus problem, of which most are based on passivity theory and internal model principle, e.g.  
\cite{Chopra:2006,Chopra:2012,Stan:2007,Munz:2011,Persis:2014,Priscoli:2015}, just to name a few. 
Consensus designs with linear and nonlinear output couplings were introduced in \cite{Chopra:2006} for heterogeneous SISO affine nonlinear MASs under the assumption of passive dynamics of agents. 
This work was then extended in \cite{Chopra:2012} where the balanced condition of inter-agent communication graph was relaxed to be strongly connected graph. 
Robust static output-feedback consensus controllers for heterogeneous SISO affine relative-degree-two passive sector-bounded MASs in presence of communication constraints were investigated in 
\cite{Munz:2011}. In \cite{Persis:2014}, distributed output tracking consensus controllers were proposed based on internal model principle and passivity for heterogeneous MIMO affine nonlinear networks of agents with relative degree one and two. In a recent work, \cite{Priscoli:2015} proposed a method to design distributed output tracking consensus controllers for heterogeneous SISO nonlinear MASs utilizing internal model principle and nonlinear controller forms that satisfying global Lipschitz conditions. Although the passive property can be found in a wide class of nonlinear systems, this approach requires the number of inputs and outputs of a nonlinear agent to be the same and further assumptions or conditions must be satisfied. Moreover, finding the energy function or the Lyapunov function to show the passivity of nonlinear agents is not always easy. 

Some other researches on consensus of nonlinear MASs consider specific problems, e.g. \cite{Yu-Chen-Cao:2011,Ding:2013}. A special class of homogeneous leader-follower MIMO affine MASs was studied in 
\cite{Yu-Chen-Cao:2011} and sufficient conditions for static consensus controllers were given based on Lipschitz assumption for agents and Lyapunov theory. 
Tracking consensus controller based on internal model principle and a specific, complicated selection of control input were introduced in \cite{Ding:2013} for a very special class of heterogeneous SISO relative-degree-one nonlinear MASs. These results are quite difficult to extend to more general contexts. 

\textcolor{blue}{ 
On the other hand, our current article proposes a systematic framework to design 
{\it linear dynamic output consensus controllers for leaderless heterogeneous MIMO general nonlinear MASs}, following the idea of designing low-rank consensus controller with non-conservative coupling strength in our previous work \cite{Nguyen:2015TAC1}. It is worth noting that there are essential differences between the current paper and \cite{Nguyen:2015TAC1} as follows. 
First, the current article deals with {\it heterogeneous nonlinear} MASs while \cite{Nguyen:2015TAC1} considered {\it homogeneous linear} MASs. 
Second, this paper proposes consensus designs for MASs with {\it directed and switching} communication topologies, but \cite{Nguyen:2015TAC1} developed consensus controllers for MASs with 
{\it undirected and fixed} structures. 
Third, the consensus controllers proposed in this paper can be freely designed to have rank $1$, however the ones in \cite{Nguyen:2015TAC1} could not. }

\textcolor{blue}{ 
The remarkable features of our proposed framework are as follows. 
First, it gives us a distributed {\it dynamic} output consensus controller design for {\it heterogeneous MIMO general nonlinear MASs with arbitrary vector relative degree} that:   
(i) has an {\it arbitrarily small but positive} coupling strength; (ii) has  {\it minimum rank}, i.e., {\it rank-$1$}. 
Second, the communication topology among agents is {\it directed} and can be {\it randomly switching} of which the component graphs {\it need not to be balanced}. 
To the best of our knowledge, there have not been similar results in the literature so far, and thus the aforementioned properties clearly show the contributions of this paper. 
}


\section{Preliminaries}
\label{pre}

\subsection{Notations and Symbols}
\label{notation}

The following notations and symbols will be used in the paper. $\mathbb{R}$, $\mathbb{R}_{-}$, and $\mathbb{C}$ stand for the sets of real, non-positive real, 
and complex number. $\mathrm{Re}(x)$ denotes the real part of a complex number $x$. 
Moreover, $\mathbf{1}_n$ and $\mathbf{0}_n$ denote the $n\times1$ vector with all elements equal to $1$ and $0$, respectively; and $I_n$ denotes the $n\times n$ identity matrix.
Next, $L_fh(x)\triangleq (\partial h(x)/\partial x)f(x)$ represents the notation for Lie derivative, and $\otimes$ stands for the Kronecker product. 
On the other hand, $\lambda(A)$ and $\lambda_{\min}(A)$ denotes the eigenvalue set and the eigenvalue with smallest, non-zero real part of $A$, respectively. 
In addition, $\succ$ and $\succeq$ denote the positive definiteness and positive semi-definiteness of a matrix. 
\textcolor{blue}{
Lastly, $\mathcal{K}_{\infty}$ denotes the class of scalar function $\gamma(x): \mathbb{R}_{+} \rightarrow \mathbb{R}_{+}$ which is continuous, strictly
increasing, unbounded, and $\gamma(0)=0$; and $\mathcal{KL}$ denotes the class of scalar function $\gamma(x,t): \mathbb{R}_{+}\times\mathbb{R}_{+} \rightarrow \mathbb{R}_{+}$ such that 
$\gamma(.,t) \in \mathcal{K}_{\infty}$ for each $t$ and $\gamma(x,t) \searrow 0$ as $t \rightarrow \infty$.
}

\subsection{Graph Theory}
\label{graph}

\textcolor{blue}{
Denote $(\mathcal{G},\mathcal{V},\mathcal{E})$ the directed graph representing the information structure in a multi-agent system composing of $N$ agents, 
where each node in $\mathcal{G}$ stands for an agent and each edge in $\mathcal{G}$ represents the interconnection between two agents; 
$\mathcal{V}$ and $\mathcal{E}$ represent the set of vertices and edges of $\mathcal{G}$, respectively. There is an edge $e_{ij} \in \mathcal{E}$ if agent $i$ receives information from agent $j$. 
The neighboring set of a vertex $i$ is denoted by $\mathcal{N}_{i} \triangleq \{j:e_{ij} \in \mathcal{E}\}$.  
Moreover, let $a_{ij}$ be elements of the adjacency matrix $\mathcal{A}$ of $\mathcal{G}$, i.e., $a_{ij}>0$ if $e_{ij} \in \mathcal{E}$ and  $a_{ij}=0$ if $e_{ij} \notin \mathcal{E}$. 
The in-degree of a vertex $i$ is denoted by $\mathrm{deg}^{\rm in}_{i} \triangleq \sum_{j=1}^{N}{a_{ij}}$, then the in-degree matrix of $\mathcal{G}$ is denoted by 
$\mathcal{D}=\mathrm{diag}\{\mathrm{deg}^{\rm in}_{i}\}_{i=1,\ldots,N}$. 
Consequently, the Laplacian matrix $\mathcal{L}$ associated to $\mathcal{G}$ is defined by $\mathcal{L}=\mathcal{D}-\mathcal{A}$. 
The out-degree of a vertex $i$ is denoted by $\mathrm{deg}^{\rm out}_{i} \triangleq \sum_{j=1}^{N}{a_{ji}}$. 
Then $\mathcal{G}$ is said to be balanced if $\mathrm{deg}^{\rm in}_{i}=\mathrm{deg}^{\rm out}_{i} \;\forall\; i=1,\ldots,N.$
}

\textcolor{blue}{
A directed path connecting vertices $i$ and $j$ in $\mathcal{G}$ is a set of consecutive edges starting from $i$ and stopping at $j$. 
Then $\mathcal{G}$ is said to have a spanning tree if there exists a node called root node from which there are directed paths to every other node. 
}

\begin{lem} \cite{Ren:2005}
\textcolor{blue}{
The Laplacian matrix $\mathcal{L}$ always has a zero eigenvalue with associated eigenvector $\mathbf{1}_{N}$, and all non-zero eigenvalues of $\mathcal{L}$ have positive real parts. 
Furthermore, $\mathcal{L}$ has only one zero eigenvalue if and only if $\mathcal{G}$ has a spanning tree.}
\end{lem}

\textcolor{blue}{
If the communication topology among agents is varied with time then we will write the time-varying terms with the time index $t$, e.g., $\mathcal{N}_{i}(t), a_{ij}(t), \mathcal{G}(t), \mathcal{L}(t)$, etc.}

\subsection{\textcolor{blue}{Consensus of Linear MASs} }
\label{linear-MAS}

\textcolor{blue}{
Let us consider an MAS composing of $N$ identical linear agents whose dynamics is described by
\begin{equation}
	\label{lin-agent}
			\dot{x}_{i} = Ax_{i}+Bu_{i}, i=1,\ldots,N,
\end{equation}
where $x_{i}\in\mathbb{R}^{n},u_{i}\in\mathbb{R}^{m}$, $A\in\mathbb{R}^{n\times n},B\in\mathbb{R}^{n\times m}$. A common consensus protocol for (\ref{lin-agent}) is
\begin{equation}
	\label{lin-consensus-ctlr}
	u_{i} = -\mu K \sum_{j\in\mathcal{N}_{i}}{a_{ij}(x_{i}-x_{j})}, i=1,\ldots,N,
\end{equation}
where $K\in\mathbb{R}^{m\times n}$ is the consensus controller gain matrix, $\mu$ is the coupling strength. 
}

\begin{lem} \cite{Z.Li:2010,Ma:2010}
\label{iff-cond}
\textcolor{blue}{
A necessary and sufficient condition for the MAS with agent dynamics (\ref{lin-agent}) to reach consensus defined as follows, 
$$\displaystyle \lim_{t \rightarrow \infty}\|x_i(t)-x_j(t)\| = 0 ~\forall~i,j=1,\ldots,N,$$ 
by the control law (\ref{lin-consensus-ctlr}) is that its communication graph $\mathcal{G}$ has a spanning tree and $A-\mu\lambda_{k}BK$ are stable for all $k=2,\ldots,N,$ where $\lambda_{k}$ are non-zero eigenvalues of $\mathcal{L}$. 
}
\end{lem}

\textcolor{blue}{
The following proposition shows a consensus design for linear MASs with non-conservative coupling strength, which serves as a basis for consensus design of nonlinear MASs with non-conservative coupling strength in the next sections. }
\begin{prop}
\label{lin-consensus}
\textcolor{blue}{
Suppose that the following conditions are satisfied: (i) the directed graph $\mathcal{G}$ representing the communication structure in the MAS (\ref{lin-agent}) has a spanning tree; (ii) $(A,B)$ is controllable; (iii) $\lambda(A) \in \mathbb{R}_{-}$. 
Then this MAS reaches consensus by the controller (\ref{lin-consensus-ctlr}) for any $\mu>0$ and $K=RB^TP$, where $R \in \mathbb{R}^{m\times m}$, $R \succ 0$, and $P\in \mathbb{R}^{n\times n}$, $P\succ 0$ is the unique solution of the following Riccati equation,
\begin{equation}
	PA+A^TP+Q-PBRB^TP=0,
\end{equation}
in which $Q\in \mathbb{R}^{n\times n}$, $Q \succeq 0$, and $(Q^{1/2},A)$ is observable. 
}
\end{prop}

\begin{IEEEproof}
\textcolor{blue}{
Based on the result of Lemma \ref{iff-cond}, the MAS (\ref{lin-agent}) will reach consensus if condition (i) is satisfied and $A-\mu\lambda_{k}BRB^TP$ are stable for all $k=2,\ldots,N,$ where $\lambda_{k}$ are non-zero eigenvalues of $\mathcal{L}$. Note that $K=RB^TP$ is in fact an LQR controller gain, and from optimal control theory \cite{Anderson:1990}, it is known that all eigenvalues of $A-BRB^TP$ are shifted to the left of the imaginary axis. On the other hand, $\mathrm{Re}(\lambda_{k})>0 \; \forall \; k=2,\ldots,N$ since $\mathcal{G}$ has a spanning tree \cite{Z.Li:2010,Ma:2010}. 
Therefore, by scaling with a scalar parameter $\mu\lambda_{k}$ with positive real part for all $k=2,\ldots,N$, the controller gain $\mu\lambda_{k}RB^TP$ still shifts all eigenvalues of $A$ to the left though it could be more or less depending on whether $\mu\lambda_{k}>1$ or $\mu\lambda_{k}<1$. Since we have assumed that $\lambda(A) \in \mathbb{R}_{-}$, this means all eigenvalues of 
$A-\mu\lambda_{k}BRB^TP$ belong to the open left half complex plane for all $k=2,\ldots,N$, and thus the consensus is achieved in the MAS (\ref{lin-agent}).  
}
\end{IEEEproof}


\section{Output Consensus of Heterogeneous SISO Nonlinear MASs with Fixed Directed Topology}
\label{consensus-siso-nonlin}

In this section, we present a novel approach to design distributed controller for output consensus problem in heterogeneous SISO nonlinear MASs with fixed topology. 
The SISO affine nonlinear MASs will be investigated first in Section \ref{siso-aff} then SISO general nonlinear MASs will consequently be studied in Section \ref{siso-gen} based on the results 
obtained for affine ones.

\subsection{Heterogeneous SISO Affine Nonlinear MASs}
\label{siso-aff}
 
Consider a network of $N$ heterogeneous SISO affine nonlinear agents whose models are described as follows,
\begin{equation}
	\label{siso-nonlin-agent}
	\begin{aligned}
		\dot{x}_i &= f_i(x_i)+g_i(x_i)u_i, \\
		y_i &= h_i(x_i), \; i=1,\ldots,N,
	\end{aligned}
\end{equation}
where $x_i \in \mathbb{R}^{n_i}$, $u_i \in \mathbb{R}$, and $y_i \in \mathbb{R}$ are the state vector, input, and output 
of the $i$th agent, respectively;  
$f_i, g_i \in \mathbb{R}^{n_i}$ and $h_i \in \mathbb{R}$ are vector-valued and scalar-valued of continuous, differentiable nonlinear functions. 

\begin{defi}
\label{rel-deg-defi}
The affine nonlinear agent (\ref{siso-nonlin-agent}) is said to have relative degree $r_{i}>0$ if 
\begin{equation}
	\label{relative-deg}
	\begin{aligned}
		L_{g_{i}}L_{f_{i}}^{k}h_{i}(x_i) &= 0 ~ {\rm as} ~ k=0,\ldots,r_{i}-2, \\
		L_{g_{i}}L_{f_{i}}^{k}h_{i}(x_i) &\neq 0 ~ {\rm as} ~ k=r_{i}-1.
	\end{aligned}
\end{equation}
\end{defi}

\begin{defi}
A multi-agent system with dynamics of agents described by (\ref{siso-nonlin-agent}) is said to reach an output consensus if 
\begin{equation}
	\displaystyle \lim_{t \rightarrow \infty}|y_i(t)-y_j(t)| = 0 ~\forall~i,j=1,\ldots,N.
\end{equation}
\end{defi}

\textcolor{blue}{
The control design problem is to find a distributed control strategy for the agents (\ref{siso-nonlin-agent}) such that their outputs cooperatively reach consensus  
while they unidirectionally exchange information through a directed graph $\mathcal{G}$. Throughout this section, we utilize the following assumption of $\mathcal{G}$. 
\begin{itemize}
	\item[{\bf A1:}] The directed graph $\mathcal{G}$ is time-invariant and has a spanning tree.
\end{itemize}
}

\textcolor{blue}{
\begin{rem}
In some practical situations, the directed graph $\mathcal{G}$ could be time-varying due to the link failures, packet losses, etc. 
This phenomenon of varied topology is usually modeled in the literature as deterministic switches (e.g., \cite{Kim:2013}, \cite{Priscoli:2015}) 
or random switches (e.g., \cite{You:2013}, \cite{Z.Li:2015}).   
For the clarity of approach representation, we first employ assumption A1 and will investigate the scenario of switching topologies later in a separated section. 
\end{rem}
}

Consequently, we employ the input-output feedback linearization method \cite{Isidori:1982} to derive linearized models of agents and accordingly convert the output consensus problem of initial nonlinear MAS to a state consensus problem of a new linearized MAS.   
More specifically, the nonlinear models of agents are changed by a diffeomorphism 
$\Phi_{i}(x_i)=[\xi_{i}^T,\eta_{i}^T]^T$ to normal forms  
\begin{equation}
	\label{siso-linearized}
	\begin{aligned}
		\dot{\xi}_{i,1} &= \xi_{i,2}, \\
		\dot{\xi}_{i,2} &= \xi_{i,3}, \\
		 & \,~ \vdots  \\
		\dot{\xi}_{i,r_{i}} &= \alpha_{i}(\xi_{i},\eta_{i})+\beta_{i}(\xi_{i},\eta_{i})u_{i}, \\
		\dot{\eta}_{i} &= \vartheta_{i}(\xi_{i},\eta_{i}), \\
		y_{i} &= \xi_{i,1}, i=1,\ldots,N,
	\end{aligned}
\end{equation} 
where $\xi_{i,k} \triangleq \Phi_{i,k}(x_i) \triangleq L_{f_{i}}^{k-1}h_{i}(x),k=1,\ldots,r_{i}$, 
$\xi_{i}=[\xi_{i,1},\ldots,\xi_{i,r_{i}}]^T \in\mathbb{R}^{r_{i}}$, $\eta_{i}\in\mathbb{R}^{n_i-r_{i}}$; 
$\alpha_{i}(\xi_{i},\eta_{i}) \triangleq L_{f_{i}}^{r_{i}}h_{i}(x)$, $\beta_{i}(\xi_{i},\eta_{i},d_{i}) \triangleq L_{g_{i}}L_{f_{i}}^{r_{i}-1}h_{i}(x)$, 
$\vartheta_{i}(\xi_{i},\eta_{i})\in\mathbb{R}^{n_i-r_{i}}$. 
\textcolor{blue}{
To avoid the finite-escape-time (FET) phenomenon and guarantee the internal stability of the closed-loop system, we employ the following assumption \cite{Sontag06},
\begin{itemize}
	\item[{\bf A2:}] The internal dynamics $\dot{\eta}_{i} = \vartheta_{i}(\xi_{i},\eta_{i})$ is input-to-state stable (ISS), i.e., there exist some functions 
	$\gamma_{i,1} \in \mathcal{KL}$ and $\gamma_{i,2} \in \mathcal{K}_{\infty}$ such that 
	\begin{equation*}
		\|\eta_{i}(t)\|_{2} \leq \gamma_{i,1}(\|\eta_{i}(0)\|_{2},t) + \gamma_{i,2}(\|\xi_{i}(t)\|_{\infty}).
	\end{equation*} 
\end{itemize}
}

Then the design problem becomes finding a control law for the linearized multi-agent system (\ref{siso-linearized}) such that linearized states $\xi_{i,1}$ of agents are consensus. 
In light of assumption A2, we are able to set the control input for the $i$th agent as follows,
\begin{equation}
	\label{nonlin-input}
		u_{i} = \frac{1}{\beta_{i}(\xi_{i},\eta_{i})}[-\alpha_{i}(\xi_{i},\eta_{i})+\hat{u}_{i}],  i=1,\ldots,N,
\end{equation}
where $\hat{u}_{i}\in \mathbb{R}$ is a new control input for the linearized subsystem. 
Since the dynamics of linearized subsystems are different, a static consensus controller cannot be derived. Instead, we will propose a dynamic controller which is able to make 
$\xi_{i,1}, i=1,\ldots,N$ converge to a common value, i.e., output consensus for (\ref{siso-nonlin-agent}) is achieved. 
Define 
\begin{equation}
	\label{rel-deg-min-s}
	r = \max_{i=1,\ldots,N} r_{i}, i=1,\ldots,N.
\end{equation}
Consequently, each agent is equipped with the following dynamic controller,
\begin{equation}
	\label{consensus-ctlr-siso-aff}
	\begin{aligned}
			\dot{\phi}_{i} &= D_{i}\xi_{i}+E_{i}\phi_{i}+G_{i}v_{i}, \\
			u_{i} &= H_{i}\phi_{i}-\frac{\alpha_{i}(\xi_{i},\eta_{i})}{\beta_{i}(\xi_{i},\eta_{i})}, i=1,\ldots,N,	
	\end{aligned}	
\end{equation}
where $\phi_{i} \in \mathbb{R}^{r-r_{i}}$ is the controller's state vector in which $\phi_{i,1}=\hat{u}_{i}$, $v_{i}\in \mathbb{R}$ is a new control input, and 
\begin{equation*}
	\begin{aligned}		
		G_{i} &= \begin{bmatrix} 0 & 0 &\ \cdots & 0 & 1 \end{bmatrix}^T \in \mathbb{R}^{r-r_{i}}, \\
		H_{i} &= \begin{bmatrix} 1/\beta_{i}(\xi_{i},\eta_{i}) & 0 & \cdots & 0 \end{bmatrix} \in \mathbb{R}^{1\times (r-r_{i})},
	\end{aligned}
\end{equation*} 
\begin{equation*}
	\label{ctlr-matrices-siso}
	\begin{aligned}
		D_{i} &=
		\begin{bmatrix}
			0 & 0 & 0 & \cdots & 0 \\
			0 & 0 & 0 & \cdots & 0 \\
			\vdots & \vdots & \vdots & & \vdots \\
			0 & 0 & 0 & \cdots & 0 \\
			0 & -b_{2} & -b_{3} & \cdots & -b_{r_{i}}
		\end{bmatrix} \in \mathbb{R}^{(r-r_{i})\times r_{i}}, \\	
		E_{i} &=
		\begin{bmatrix}
			0 & 1 & \cdots & 0 \\
			0 & 0 & \cdots & 0 \\
			\vdots & \vdots & & \vdots \\
			0 & 0 & \cdots & 1 \\
			-b_{r_{i}+1} & -b_{r_{i}+2} & \cdots & -b_{r}
		\end{bmatrix} \in \mathbb{R}^{(r-r_{i})\times (r-r_{i})}, \\
	\end{aligned}
\end{equation*} 
of which $b_{2},\ldots,b_{r}$ are coefficients of the following characteristic equation whose poles are in $\mathbb{R}_{-}$,
\begin{equation}
	\label{charac-eq}
	s^{r}+b_{r}s^{r-1}+\cdots+b_{2}s=0. 
\end{equation}
Note that the free coefficient is chosen to be $0$ to ensure a non-zero consensus. 
Denote $\hat{\xi}_{i}=[\xi_{i}^T,\phi_{i}^T]^T$, then the overall linearized dynamics of agents are made identical by the dynamic controllers (\ref{consensus-ctlr-siso-aff}), and has the following representation,
\begin{equation}
	\label{subsys}
	\begin{aligned}
			\dot{\hat{\xi}}_{i} &= A\hat{\xi}_i+Bv_i, \\
			u_{i} &= \begin{bmatrix} 0 & H_{i} \end{bmatrix}\hat{\xi}_{i}-\alpha_{i}(\xi_{i},\eta_{i})/\beta_{i}(\xi_{i},\eta_{i}), i=1,\ldots,N,	
	\end{aligned}		 
\end{equation} 
where   
\begin{equation*}
	\label{AB-eq}
	A=
	\begin{bmatrix}
		0 & 1 & 0 & \cdots & 0 \\
		0 & 0 & 1 & \cdots & 0 \\
		\vdots & \vdots & \vdots & & \vdots \\
		0 & 0 & 0 & \cdots & 1 \\
		0 & -b_{2} & -b_{3} & \cdots & -b_{r}
	\end{bmatrix} \in \mathbb{R}^{r\times r},
	B=\begin{bmatrix} 0 \\ 0 \\ \cdots \\ 0 \\ 1 \end{bmatrix} \in \mathbb{R}^{r}. 
\end{equation*} 

%

%

We are now ready to state a foundation result of this paper in the following theorem where the coupling strength in the consensus law for nonlinear MASs is {\it non-conservative}. 

\begin{thm}
\label{siso-aff-thm}
The heterogeneous SISO affine nonlinear MAS (\ref{siso-nonlin-agent}) reaches an output consensus by the local dynamic controllers (\ref{consensus-ctlr-siso-aff}) and the cooperative controls   
\begin{equation}
	\label{lin-ctlr-siso-aff}
	v_{i} = -\mu (\hat{r}B^TP_1)\sum_{j\in\mathcal{N}_i}a_{ij}(\hat{\xi}_{i}(t)-\hat{\xi}_{j}(t)), 
\end{equation}
for any $\mu > 0$, where $\hat{r} > 0$ and $P_1 \in \mathbb{R}^{r\times r}$ is the unique positive definite solution of the following Riccati equation
\begin{equation}
	\label{Req-local}
	P_1A+A^TP_1+Q_1-\hat{r}P_1BB^TP_1=0,
\end{equation}
of which $Q_1 \in \mathbb{R}^{r\times r}$, $Q_1 \succeq 0$, and $(Q_{1}^{1/2},A)$ is observable.
\end{thm}

\begin{IEEEproof}
\textcolor{blue}{
Since the incorporated models of linearized agents in (\ref{subsys}) are homogeneous, linear, and all of their poles are in $\mathbb{R}_{-}$, we can immediately apply the result 
of Proposition \ref{lin-consensus} in Section \ref{linear-MAS} for designing a distributed consensus controller for (\ref{subsys}) under the form of (\ref{lin-ctlr-siso-aff}). 
Note that in the current situation each linearized agent is SISO, so the weighting matrix $R$ becomes a scalar parameter that we denoted by $\hat{r}$. 
Consequently, in combination with the local dynamic controllers (\ref{consensus-ctlr-siso-aff}), it gives us the output consensus of the initial nonlinear MAS (\ref{siso-nonlin-agent}).  
}
\end{IEEEproof}

\begin{figure}[ht!]
		\centering
		\psfrag{u}{\Huge $u_i$}
		\psfrag{v}{\Huge $v_{i}$}
		\psfrag{xi}{\Huge $\xi_{i}$}
		\psfrag{xii}{\Huge $\hat{\xi}_{i}$}
		\psfrag{xij}{\Huge $\hat{\xi}_{j}$}
		\psfrag{K}{\Huge $\displaystyle \mu (\hat{r}B^TP_1)\sum_{(i,j) \in {\cal E}}{a_{ij}\hat{\xi}_{j}}$}
		\psfrag{y}{\Huge $y_i$}
		\psfrag{d}{\Huge $\displaystyle -\frac{\alpha_{i}}{\beta_{i}}$}
		\psfrag{phi}{\Huge $\Phi_{i}(x_i)$}
		\psfrag{p}{\scalebox{1.3}{\huge Agent $i$th}}
		\psfrag{int}{\Huge $C(s)$}
		\psfrag{coop}{\Huge \it \color{blue}{Cooperative control}}
		\psfrag{dyn}{\Huge \it \color{red}{Local dynamic controller}}
		\scalebox{0.3}{\includegraphics{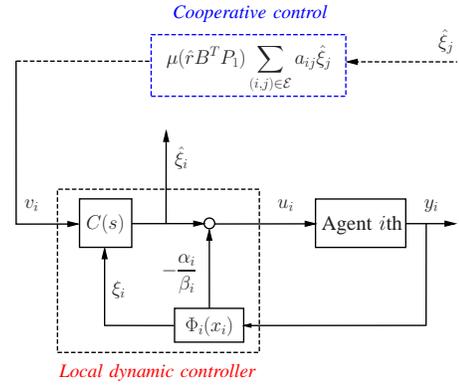}}
		\caption{Block diagram of distributed output consensus design for SISO affine nonlinear MASs based on input-output feedback linearization.}
		\label{siso-aff-diag}
	\end{figure}

The control design for the whole system is demonstrated in Figure \ref{siso-aff-diag}, where $C(s)$ represents the transfer function of identical linearized systems (\ref{subsys}). 

\begin{rem}
\textcolor{blue}{
It can be seen in Theorem \ref{siso-aff-thm} that $\mu$ can be arbitrarily chosen as long as it is positive. 
On the other hand, in other researches, e.g., \cite{Z.Li:2010,Ma:2010,F.Xiao:2007,Movric:2014}, $\mu$ is lower bounded by $1/\mathrm{Re}(\lambda_{\min}(\mathcal{L}))$ which can be extremely big as the number of agents increases and the inter-agent communication topology is sparse and hence is very conservative. Moreover, $\lambda_{\min}(\mathcal{L})$ is a global information and therefore to make the consensus law fully distributed, other methods need to be further developed to estimate this global term, e.g., adaptive designs \cite{Z.Li:2015}. This unexpectedly increases the complexity of the control design and implementation. Nevertheless, this conservatism is removed in our work, which makes our consensus design non-conservative and more effective in design and implementation. }
\end{rem}

\begin{rem}
\textcolor{blue}{
The output consensus design in Theorem \ref{siso-aff-thm} relies on the output $y_{i}$, its first-order and higher-order derivatives which may not be available in some practical systems. 
In that cases, local estimation techniques 
can be employed to obtain the approximated values of those unmeasurable derivatives. 
Let us denote $\theta_{i}=C\hat{\xi}_{i}$ the partial information that could be exchanged among agents with $C \in \mathbb{R}^{1\times r}$, then we can employ a 
decentralized Luenberger observer \cite{Nguyen:2015j} for each agent as follows,
\begin{equation}
	\label{obs-1}
	\begin{array}{ll}
		\dot{\check{\xi}}_{i} &= A\check{\xi}_{i}+Bv_{i}+M(\theta_{i}-\check{\theta}_{i}), \\
		\check{\theta}_{i} &= C\check{\xi}_{i},i=1,\ldots,N,
	\end{array}
\end{equation} 
where $M \in \mathbb{R}^{r}$. 
Denote $e_{i}=\hat{\xi}_{i}-\check{\xi}_{i}$ the error vector between the real state $\hat{\xi}_{i}$ and the estimated state $\check{\xi}_{i}$.  
Then by subtracting (\ref{subsys}) with (\ref{obs-1}), we obtain the following error model,
\begin{equation}
	\label{error}
		\dot{e}_{i} = (A-MC)e_{i}.
\end{equation}
As a result, by selecting the observer gain $M$ such that $A-MC$ is stable, $e(t) \rightarrow 0$ as $t \rightarrow \infty$, i.e., $\check{\xi}_{i} \rightarrow \hat{\xi}_{i}$ as $t \rightarrow \infty$. 
Lastly, the cooperative control input $v_{i}$ is modified by
\begin{equation}
	\label{output-ctlr}
		v_{i} = -\mu (\hat{r}B^TP_1)\sum_{j\in\mathcal{N}_i}a_{ij}(\check{\xi}_{i}(t)-\check{\xi}_{j}(t)),
\end{equation}
where $\check{\xi}$ is obtained from the local observer (\ref{obs-1}).  
}
\end{rem}


\textcolor{blue}{
In Theorem \ref{siso-aff-thm}, a local rank-$r$ Riccati equation (\ref{Req-local}) needs to be solved to obtain the consensus controller, which would cost more computational time than expected for high relative degree nonlinear agents. Hence, we next propose a method to derive a {\it minimum rank} distributed consensus controller by solving a local rank-$1$, i.e., scalar Riccati equation. The controller is therefore fully analytical, which requires no additional time for solving Riccati equation. }

First, we choose $b_{2},\ldots,b_{r}$ such that matrix $A$ defined in (\ref{AB-eq}) has only one eigenvalue at the origin while other eigenvalues belong to the open left half complex plane. 
Let $\nu^T\in \mathbb{R}^{1\times r}$ be the left eigenvector of $A$ associated with the eigenvalue $0$. Second, select 
$Q_1=\nu q_1\nu^T$ where $q_1 > 0$. 
Suppose that $(Q_1^{1/2},A)$ is still observable. Then, the rank-$1$ distributed dynamic consensus controller is derived as follows. 

\begin{thm}
\label{siso-aff-coro}
The heterogeneous SISO affine nonlinear MAS (\ref{siso-nonlin-agent}) reaches an output consensus by the distributed dynamic consensus controllers (\ref{consensus-ctlr-siso-aff}) with  
the rank-$1$ cooperative inputs $v_{i},i=1,\ldots,N$, synthesized as follows,
\begin{equation}
	\label{lin-ctlr-siso-aff-rank1}
	v_{i} = -\mu \sqrt{q_1\hat{r}}\nu^T \sum_{j\in\mathcal{N}_i}a_{ij}(\hat{\xi}_{i}(t)-\hat{\xi}_{j}(t)). 
\end{equation}
Furthermore, the consensus speed, i.e., the smallest non-zero absolute of real parts of closed-loop eigenvalues, is equal to
\begin{equation}
	\label{consensus-speed-aff}
	\min\left\{ \mu\sqrt{q_1\hat{r}}B^T\nu\mathrm{Re}(\lambda_{\min}(L)),\lambda_{\min}(-A) \right\}.
\end{equation}
\end{thm}

\begin{IEEEproof}
\textcolor{blue}{ 
First, we prove the rank-$1$ consensus controller's formula. Let $P_{1}=\nu p_{1} \nu^{T}, p_1> 0$, then substituting $Q_{1}$ and $P_{1}$ back to the Riccati equation (\ref{Req-local}), we obtain
\begin{equation}
	\nu \left(p_{1}\nu^{T}A+A^T\nu p_1+q_{1}-\hat{r}\nu^TBB^T\nu p_{1}^{2} \right)\nu^{T} = 0,
\end{equation} 
which is equivalent to the vanishment of the expression inside the bracket. Since $\nu^{T}A=0$, this implies
$q_{1}-\hat{r}\nu^TBB^T\nu p_{1}^{2}=0,$ 
which leads to $p_{1}=\sqrt{q_{1}}/\sqrt{\hat{r}}B^T\nu.$ 
Since (\ref{Req-local}) has a unique positive semidefinite solution, 
$P_1=\nu \sqrt{q_{1}}/(\sqrt{\hat{r}}B^T\nu) \nu^{T}$ is indeed that unique one. 
Consequently, substituting this value of $P_{1}$ into the cooperative control input (\ref{lin-ctlr-siso-aff}) gives us (\ref{lin-ctlr-siso-aff-rank1}). 
Obviously, $\mathrm{rank}(v_{i})=1$, so together with (\ref{consensus-ctlr-siso-aff}) we derive a rank-$1$ distributed dynamic consensus controller. 
}

\textcolor{blue}{ 
Next, we reveal how to obtain the consensus speed. Employing the same process in the proof of Theorem 3 in \cite{Nguyen:2015TAC1}, we can easily show that the eigenvalue set of the closed-loop dynamics of the linearized MAS is given by
\begin{equation}
	\left(\bigcup_{\gamma \in \lambda(L),\gamma \neq 0}-\mu\gamma\sqrt{q_1\hat{r}}B^T\nu\right)\bigcup\left(\lambda(A)\backslash \{0\}\right).
\end{equation} 
Thus, the consensus speed  is determined by (\ref{consensus-speed-aff}). 
}
\end{IEEEproof}

\begin{rem}
\textcolor{blue}{
It can be observed from Theorem \ref{siso-aff-coro} that $\mu$, the pole of $A$ closest to the imaginary axis, and $q_{1}$ and $\hat{r}$ are parameters that affect to the consensus speed. Hence, we may adjust them to obtain an expected consensus speed. 
}
\end{rem}

\subsection{Heterogeneous SISO General Nonlinear MASs}
\label{siso-gen}

In this scenario, the models of agents are in the following general form
\begin{equation}
	\label{siso-nonlin-gen}
	\begin{aligned}
		\dot{x}_i &= f_i(x_i,u_i), \\
		y_i &= h_i(x_i), \; i=1,\ldots,N,
	\end{aligned}
\end{equation}
where $x_i \in \mathbb{R}^{n_i}$, $u_i \in \mathbb{R}$, and $y_i \in \mathbb{R}$ are the state vector, input, and output 
of the $i$th agent, respectively;  
$f_i \in \mathbb{R}^{n_i}$ and $h_i \in \mathbb{R}$ are vector-valued and scalar-valued of continuous, differentiable nonlinear functions. 

\begin{defi}
	The general nonlinear agent (\ref{siso-nonlin-gen}) is said to have relative degree $r_{i}$ if 
	\begin{equation}
		\label{relative-deg-g}
		\begin{aligned}
			\frac{\partial}{\partial u_{i}} L_{f_{i}}^{k}h_{i}(x_i) &= 0 ~ {\rm as} ~ k=0,\ldots,r_{i}-1, \\
			\frac{\partial}{\partial u_{i}} L_{f_{i}}^{k}h_{i}(x_i) &\neq 0 ~ {\rm as} ~ k=r_{i}.
		\end{aligned}
	\end{equation}
\end{defi}

The dynamic distributed controller (\ref{consensus-ctlr-siso-aff}) for affine nonlinear MASs cannot be utilized in this scenario. 
However, it is possible if we consider the following {\it augmented models} of agents which are affine, 
\begin{equation}
	\label{siso-nonlin-augmented}
	\begin{aligned}
		\dot{\tilde{x}}_i &= \tilde{f}_i(\tilde{x}_i)+\tilde{g}(\tilde{x}_i)\dot{u}_i, \\
		y_i &= \tilde{h}_i(\tilde{x}_i), \; i=1,\ldots,N,
	\end{aligned}
\end{equation}
where $\tilde{x}_i \triangleq [x_i^T,u_{i}]^T$, $\tilde{f}_i(\tilde{x}_i) \triangleq [f_i(x_i,u_i)^T,0]^T$, 
$\tilde{g}_i(\tilde{x}_i) \triangleq [0_{n_i}^T,1]^T$, $\tilde{h}_i(\tilde{x}_i) \triangleq h_i(x_i)$. 
Consequently, it can be easily checked that the relative degree of the augmented affine nonlinear agents (\ref{siso-nonlin-augmented}), in the sense of Definition \ref{rel-deg-defi}, 
are $r_{i}+1, i=1,\ldots,N$. 
Similarly to the case of affine nonlinear MASs, we employ the input-output linearization feedback approach with diffeomorphisms 
$\tilde{\Phi}_{i}(\tilde{x}_i)=[\tilde{\xi}_{i}^T,\tilde{\eta}_{i}^T]^T$ \textcolor{blue}{assumed that the internal dynamics $\tilde{\eta}_{i}$ is ISS.} Then the linearized subsystems of agents are obtained by the following control inputs
\begin{equation}
	\label{nonlin-input-1}
		\dot{u}_i = \frac{1}{\tilde{\beta}_{i}(\tilde{\xi}_{i},\tilde{\eta}_{i})}[-\tilde{\alpha}_{i}(\tilde{\xi}_{i},\tilde{\eta}_{i})+\tilde{u}_{i}], i=1,\ldots,N,
\end{equation}
where $\tilde{\beta}_{i}(\tilde{\xi}_{i},\tilde{\eta}_{i}) \triangleq L_{\tilde{g}_{i}}L_{\tilde{f}_{i}}^{r_{i}}\tilde{h}_{i}(\tilde{x}_i)$; 
$\tilde{\alpha}_{i}(\tilde{\xi}_{i},\tilde{\eta}_{i}) \triangleq L_{\tilde{f}_{i}}^{r_{i}+1}\tilde{h}_{i}(\tilde{x}_i)$;
$\tilde{u}_{i}\in \mathbb{R}$ is a new control input for the linearized subsystem of the augmented nonlinear agents (\ref{siso-nonlin-augmented}). 
Let $r$ be defined as in (\ref{rel-deg-min-s}). 
Then we propose the following dynamic controller
\begin{equation}
	\label{consensus-ctlr-siso-gen}
	\begin{aligned}
			\dot{\tilde{\phi}}_{i} &= \tilde{D}_{i}\tilde{\xi}_{i}+\tilde{E}_{i}\tilde{\phi}_{i}+\tilde{G}_{i}\tilde{v}_{i}, \\
			w_{i} &= \tilde{H}_{i}\tilde{\phi}_{i}-\frac{\tilde{\alpha}_{i}(\tilde{\xi}_{i},\tilde{\eta}_{i})}{\tilde{\beta}_{i}(\tilde{\xi}_{i},\tilde{\eta}_{i})}, \\
			\dot{u}_{i} &= w_{i}, i=1,\ldots,N,	
	\end{aligned}	
\end{equation}
where $\tilde{\phi}_{i} \in \mathbb{R}^{r-r_{i}}$ is a vector of controller's states in which $\tilde{\phi}_{i,1}=\tilde{u}_{i}$; $\tilde{v}_{i}\in \mathbb{R}$ is a new control input; 
$w_{i}\in \mathbb{R}$ is an additional state of the controller; $\tilde{E}_{i}$ and $\tilde{G}_{i}$ are defined as follows, 
\begin{equation*}
	\begin{aligned}
		\tilde{D}_{i} &=
		\begin{bmatrix}
			0 & 0 & 0 & \cdots & 0 \\
			0 & 0 & 0 & \cdots & 0 \\
			\vdots & \vdots & \vdots & & \vdots \\
			0 & 0 & 0 & \cdots & 0 \\
			0 & -\tilde{b}_{2} & -\tilde{b}_{3} & \cdots & -\tilde{b}_{r_{i}+1}
		\end{bmatrix} \in \mathbb{R}^{(r-r_{i})\times(r_{i}+1)}, \\
		\tilde{E}_{i} &=
		\begin{bmatrix}
			0 & 1 & \cdots & 0 \\
			0 & 0 & \cdots & 0 \\
			\vdots & \vdots & & \vdots \\
			0 & 0 & \cdots & 1 \\
			-\tilde{b}_{r_{i}+2} & -\tilde{b}_{r_{i}+2} & \cdots & -\tilde{b}_{r+1}
		\end{bmatrix} \in \mathbb{R}^{(r-r_{i})\times (r-r_{i})},
	\end{aligned}			
\end{equation*}
\begin{equation*}
	\begin{aligned}	
		\tilde{G}_{i} &= \begin{bmatrix} 0 & 0 &\ \cdots & 0 & 1 \end{bmatrix}^T \in \mathbb{R}^{r-r_{i}}, \\
		\tilde{H}_{i} &= \begin{bmatrix} 1/\tilde{\beta}_{i}(\tilde{\xi}_{i},\tilde{\eta}_{i}) & 0 & \cdots & 0 \end{bmatrix} \in \mathbb{R}^{1\times (r-r_{i})}, 
	\end{aligned}
\end{equation*} 
of which $\tilde{b}_{2},\ldots,\tilde{b}_{r+1}$ are coefficients of the following characteristic equation whose poles are in $\mathbb{R}_{-}$,
\begin{equation}
	\label{charac-eq-1}
	s^{r+1}+\tilde{b}_{r+1}s^{r}+\cdots+\tilde{b}_{2}s=0. 
\end{equation}
This controller can be viewed as a cascade of two controllers in which the first one composes of the first two equations in (\ref{consensus-ctlr-siso-gen}) while the second one is an integrator corresponding to the last equation (\ref{consensus-ctlr-siso-gen}). 
Subsequently, the closed-loop linearized dynamics of agents are made homogeneous by the dynamic controllers (\ref{consensus-ctlr-siso-gen}), and has the following representation,
\begin{equation}
	\label{subsys-1}
	\begin{aligned}
			\dot{\zeta}_{i} &= \tilde{A}\zeta_i+\tilde{B}\tilde{v}_i, \\
			w_{i} &= \begin{bmatrix} 0 & \tilde{H}_{i} \end{bmatrix}\hat{\xi}_{i}-\tilde{\alpha}_{i}(\tilde{\xi}_{i},\tilde{\eta}_{i})/\tilde{\beta}_{i}(\tilde{\xi}_{i},\tilde{\eta}_{i}), i=1,\ldots,N,	
	\end{aligned}		 
\end{equation} 
where 
\begin{equation}
	\label{AB-eq-1}
	\begin{aligned}
		\zeta_{i} &= \begin{bmatrix} \tilde{\xi}_{i}^T & \tilde{\phi}_{i}^T \end{bmatrix}^T, \\
		\tilde{A} &=
		\begin{bmatrix}
			0 & 1 & 0 & \cdots & 0 \\
			0 & 0 & 1 & \cdots & 0 \\
			\vdots & \vdots & \vdots & & \vdots \\
			0 & 0 & 0 & \cdots & 1 \\
			0 & -\tilde{b}_{2} & -\tilde{b}_{3} & \cdots & -\tilde{b}_{r+1}
		\end{bmatrix} \in \mathbb{R}^{(r+1)\times (r+1)}, \\
		\tilde{B} &= \begin{bmatrix} 0 & 0 & \cdots & 0 & 1 \end{bmatrix}^T \in \mathbb{R}^{r+1}. 
	\end{aligned}
\end{equation}
%
%
%

\begin{figure}[ht!]
		\centering
		\psfrag{u}{\Huge $u_i$}
		\psfrag{w}{\Huge $w_i$}
		\psfrag{v}{\Huge $\tilde{v}_{i}$}
		\psfrag{xi}{\Huge $\tilde{\xi}_{i}$}
		\psfrag{zetai}{\Huge $\zeta_{i}$}
		\psfrag{zetaj}{\Huge $\zeta_{j}$}
		\psfrag{K}{\Huge $\displaystyle \mu (\tilde{r}\tilde{B}^T\tilde{P}_1)\sum_{(i,j) \in {\cal E}}{a_{ij}\zeta_{j}}$}
		\psfrag{y}{\Huge $y_i$}
		\psfrag{d}{\Huge $\displaystyle -\frac{\tilde{\alpha}_{i}}{\tilde{\beta}_{i}}$}
		\psfrag{phi}{\Huge $\tilde{\Phi}_{i}(\tilde{x}_i)$}
		\psfrag{p}{\scalebox{1.3}{\huge Agent $i$th}}
		\psfrag{int}{\Huge $\tilde{C}(s)$}
		\psfrag{int1}{\Huge $\displaystyle \frac{1}{s}$}
		\psfrag{coop}{\Huge \it \color{blue}{Cooperative control}}
		\psfrag{dyn}{\Huge \it \color{red}{Local dynamic controller}}
		\scalebox{0.3}{\includegraphics{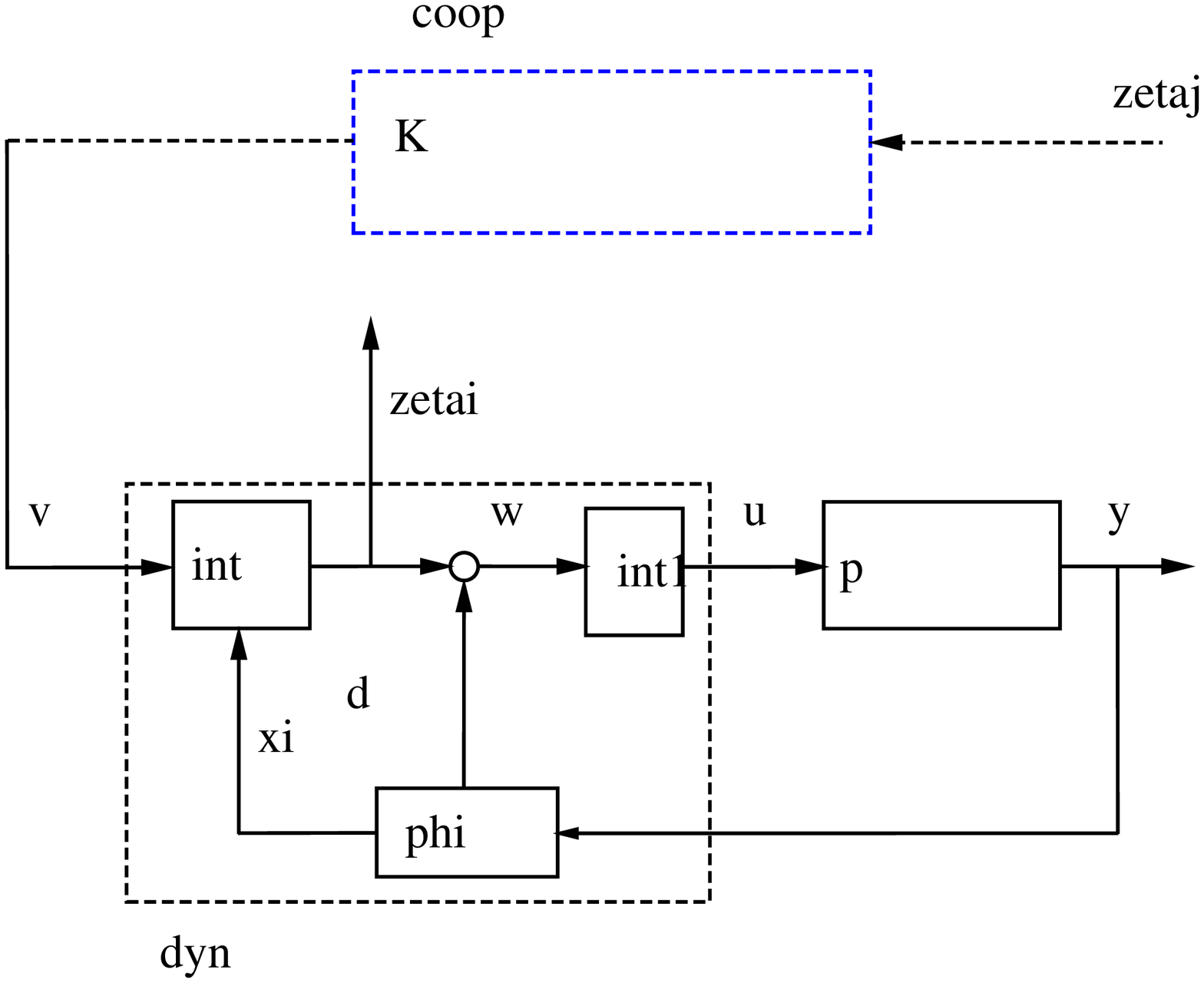}}
		\caption{Block diagram of distributed output consensus design for SISO general nonlinear MASs based on input-output feedback linearization.}
		\label{siso-gen-diag}
	\end{figure}

The control design for the whole system is demonstrated in Figure \ref{siso-gen-diag}, where $\tilde{C}(s)$ represents the transfer function of identical linearized systems in 
(\ref{subsys-1}) from the inputs $\tilde{v}_{i}$ to the outputs $w_{i}$, $i=1,\ldots,N$.  

\begin{figure*}[!t]
\normalsize
\setcounter{MYtempeqncnt}{\value{equation}}
\setcounter{equation}{33}
\begin{equation}
	\label{decoupled-matrix}
	\Pi_{i}(x_i) = \begin{bmatrix}
		L_{g_{i,1}}L_{f_{i}}^{r_{i,1}-1}h_{i,1}(x_i) & L_{g_{i,2}}L_{f_{i}}^{r_{i,1}-1}h_{i,1}(x_i) & \cdots & L_{g_{i,m_i}}L_{f_{i}}^{r_{i,1}-1}h_{i,1}(x_i) \\
		L_{g_{i,1}}L_{f_{i}}^{r_{i,2}-1}h_{i,2}(x_i) & L_{g_{i,2}}L_{f_{i}}^{r_{i,2}-1}h_{i,2}(x_i) & \cdots & L_{g_{i,m_i}}L_{f_{i}}^{r_{i,2}-1}h_{i,2}(x_i) \\
		\vdots & \vdots & \ddots & \vdots & \\
		L_{g_{i,1}}L_{f_{i}}^{r_{i,p}-1}h_{i,p}(x_i) & L_{g_{i,2}}L_{f_{i}}^{r_{i,p}-1}h_{i,p}(x_i) & \cdots & L_{g_{i,m_i}}L_{f_{i}}^{r_{i,p}-1}h_{i,p}(x_i) 
		\end{bmatrix}.
\end{equation}

\setcounter{equation}{\value{MYtempeqncnt}}
\hrulefill
\vspace*{4pt}
\end{figure*}	

Similarly to the scenario of SISO affine nonlinear MASs, we can design rank-$1$ distributed dynamic consensus controllers for the MAS (\ref{siso-nonlin-gen}). 
Let us select $\tilde{b}_{2},\ldots,\tilde{b}_{r+1}$ such that matrix $\tilde{A}$ defined in (\ref{AB-eq-1}) has only one eigenvalue at the origin while other eigenvalues belong to the open left half complex plane. 
Denote $\tilde{\nu}^T\in \mathbb{R}^{1\times(r+1)}$ the left eigenvector of $\tilde{A}$ associated with the eigenvalue $0$. Consequently, choose 
$\tilde{Q}_1=\tilde{\nu}\tilde{q}_1\tilde{\nu}^T,$ where $\tilde{q}_1 > 0$. 
Assuming that $(\tilde{Q}_{1}^{1/2},\tilde{A})$ is still observable, then the rank-$1$ distributed dynamic consensus controller in this case is introduced in the following theorem. 

\begin{thm}
\label{siso-gen-coro}
The heterogeneous SISO general nonlinear MAS (\ref{siso-nonlin-gen}) reaches an output consensus by the distributed dynamic consensus controllers (\ref{consensus-ctlr-siso-gen}) with  
the rank-$1$ cooperative inputs $\tilde{v}_{i},i=1,\ldots,N$, synthesized as follows,
\begin{equation}
	\label{lin-ctlr-siso-gen-rank1}
	\tilde{v}_{i} = -\mu \sqrt{\tilde{q}_1\tilde{r}}\tilde{\nu}^T \sum_{j\in\mathcal{N}_i}a_{ij}(\zeta_{i}(t)-\zeta_{j}(t)). 
\end{equation}
Furthermore, the consensus speed is equal to
\begin{equation}
	\label{consensus-speed-gen}
	\min\left\{ \mu\sqrt{\tilde{q}_1\tilde{r}}\tilde{B}^T\tilde{\nu}\mathrm{Re}(\lambda_{\min}(L)),\lambda_{\min}(-\tilde{A}) \right\}.
\end{equation}
\end{thm}

\begin{IEEEproof}
The proof of this theorem is the same as that of Theorem \ref{siso-aff-coro}, so we omit it for brevity.   
\end{IEEEproof}


\section{Output Consensus of Heterogeneous MIMO Nonlinear MASs with Fixed  Directed Topology}
\label{consensus-mimo-nonlin}


Consider a MAS composing of $N$ heterogeneous MIMO affine nonlinear agents whose models are described as follows,
\begin{equation}
	\label{mimo-nonlin-agent-aff}
	\begin{aligned}
		\dot{x}_i &= f_i(x_i)+G_i(x_i)u_i, \\
		y_i &= h_i(x_i), \; i=1,\ldots,N,
	\end{aligned}
\end{equation}
where $x_i \in \mathbb{R}^{n_i}$, $u_i \in \mathbb{R}^{m_i}$, and $y_i \in \mathbb{R}^{p}$ are the state, input, and output vectors
of the $i$th agent, respectively;  
$f_i(x_i) \in \mathbb{R}^{n_i}$, $h_i(x_i) \in \mathbb{R}^{p}$, and $G_i \in \mathbb{R}^{n_i\times m_i}$ are vector-valued and matrix-valued of continuous, differentiable nonlinear functions; 
$h_{i}(x_i)=[h_{i,1}(x_i),\ldots,h_{i,p}(x_i)]^T$, $h_{i,\tau}(x_i) \in \mathbb{R} \,\forall\, \tau=1,\ldots,p$; $G_i(x_i)=[g_{i,1}(x_i),\ldots,g_{i,m_i}(x_i)]$, $g_{i,j}(x_i) \in \mathbb{R}^{n_i} \,\forall\, j=1,\ldots,m_i$. 
\textcolor{blue}{The communication topology $\mathcal{G}$ among agents is also assumed to satisfy assumption A1 in Section \ref{consensus-siso-nonlin}.} 
The output dimensions of all agents are equal to be meaningful in the context of output consensus, which is defined as follows.

\begin{defi}
\label{mimo-consensus-defi}
The outputs of agents whose models are described by (\ref{mimo-nonlin-agent-aff}) are said to reach a consensus if
\begin{equation}
	\label{mimo-output-consensus}
	\lim_{t \rightarrow \infty} \|y_{i}(t)-y_{j}(t)\|=0 \; \forall \; i,j=1,\ldots,N.
\end{equation}
\end{defi}

In the use of input-output feedback linearization for nonlinear systems \cite{Isidori:1982}, the dimensions of input and output are usually assumed to be equal, however we consider here a more general context where those dimensions may be different. Accordingly, the definition of {\it vector relative degree} \cite{Isidori:1982} can be modified as follows. 
\begin{defi}
The MIMO nonlinear system (\ref{mimo-nonlin-agent-aff}) is said to have an {\it extended vector relative degree} $\{r_{i,1},r_{i,2},\ldots,r_{i,p}\}$ at a point $\hat{x}_{i} \in \mathbb{R}^{n_i}$ if the following conditions hold.
\begin{itemize}
	\item[(i)] $L_{g_{i,j}}L_{f_{i}}^{k}h_{i,l}(x_i)=0 \; \forall \; k=0,\ldots,r_{i,l}-2; j=1,\ldots,m_i$; $1 \leq l \leq p$; and for all $x_i$ in a neighborhood of $\hat{x}_{i}$. 
	\item[(ii)] ${\rm rank}(\Pi_{i}(x_i))=p$ at the point $\hat{x}_{i}$, where $\Pi_{i}(x_i)$ is defined in (\ref{decoupled-matrix}). 
\end{itemize}
\end{defi}

\textcolor{blue}{
Note that condition (ii) above is satisfied if and only if $p \leq m_{i}$, i.e., the {\it extended vector relative degree} is only defined for the MIMO affine nonlinear systems that have the number of outputs not more than the number of inputs. If $p=m_{i}$ then it reduces to the {\it vector relative degree} in \cite{Isidori:1982} since condition (ii) means $\Pi_{i}(x_i)$ is invertible at 
$\hat{x}_{i}$. Furthermore, this {\it extended vector relative degree} allows us to treat a scenario that the {\it vector relative degree} in \cite{Isidori:1982} cannot, where some agents in a MIMO affine nonlinear MASs have the same number of inputs and outputs but other agents do not.  
}

\addtocounter{equation}{1}

To design a distributed output consensus controller for the MAS (\ref{mimo-nonlin-agent-aff}), we also try to obtain linearized models of agents and then design the consensus controller based on those models.  
Similarly to the scenario of SISO nonlinear MASs, we utilize the input-output feedback linearization technique for nonlinear agents (\ref{mimo-nonlin-agent-aff}) that can be processed as follows. 
Denote $\kappa_{i} = \sum_{l=1}^{p} r_{i,l}, i=1,\ldots,N.$ 
Consequently, the agents' models are changed to normal forms by a diffeomorphism $\Phi_{i}(x_i)=[\xi_{i}^T,\eta_{i}^T]^T$ where  
\begin{equation}
	\begin{aligned}
		\Phi_{i}(x_i) \triangleq & \, [\Phi_{i,1}^{1}(x_i),\ldots,\Phi_{i,r_{i,1}}^{1}(x_i),\ldots,\Phi_{i,1}^{p}(x_i),\ldots, \\
		& \Phi_{i,r_{i,p}}^{p}(x_i),\Phi_{i,\kappa_{i}+1}(x_i),\ldots,\Phi_{i,n_i}(x_i)]^T, \\
		\xi_{i,j}^{l} \triangleq & \, \Phi_{i,j}^{l}(x_i) \triangleq L_{f_{i}}^{j-1}h_{i,l}(x_i) \\
		& \; \forall \; j=1,\ldots,r_{i,l}; \; \forall \; l=1,\ldots,p, \\
		\xi_{i} \triangleq & \, [\xi_{i,1}^{1},\ldots,\xi_{i,r_{i,1}}^{1},\ldots,\xi_{i,1}^{p},\ldots,\xi_{i,r_{i,p}}^{p}]^T, \\
		\eta_{i} \triangleq & \, [\Phi_{i,\kappa_{i}+1}(x_i),\ldots,\Phi_{i,n_i}(x_i)]^T.
	\end{aligned}
\end{equation}
The linearized model for agent $i$th is as follows,
\begin{equation}
	\label{mimo-linearized}
	\begin{aligned}
		\dot{\xi}_{i,1}^{l} &= \xi_{i,2}^{l}, \\
		\dot{\xi}_{i,2}^{l} &= \xi_{i,3}^{l}, \\
		 & \,~ \vdots  \\
		\dot{\xi}_{i,r_{i,l}}^{l} &= \alpha_{i}^{l}(\xi_{i},\eta_{i})+\sum_{k=1}^{m_i}\beta_{i,k}^{l}(\xi_{i},\eta_{i})u_{i,k}, \\
		\dot{\eta}_{i} &= \vartheta_{i}(\xi_{i},\eta_{i})+\sum_{k=1}^{m_i}\chi_{i,k}(\xi_{i},\eta_{i})u_{i,k}, \\
		y_{i}^{l} &= \xi_{i,1}^{l} \; \forall \; l=1,\ldots,p; i=1,\ldots,N,
	\end{aligned}
\end{equation} 
where $u_{i,k}$ is the $k$th control input of the $i$th agent, $k=1,\ldots,m_i$, and
\begin{equation*}
	\begin{aligned}
		\alpha_{i}^{l}(\xi_{i},\eta_{i}) &\triangleq L_{f_{i}}^{r_{i,l}}h_{i,l}(x_i),\beta_{i,k}^{l}(\xi_{i},\eta_{i}) \triangleq L_{g_{i,k}}L_{f_{i}}^{r_{i,l}-1}h_{i,l}(x_i), \\
		\vartheta_{i}(\xi_{i},\eta_{i}) &\in \mathbb{R}^{n_i-\kappa_i}, \\
		\chi_{i,k}(\xi_{i},\eta_{i}) &\triangleq [L_{g_{i,k}}\Phi_{i,\kappa_i+1}(x_i),\ldots,L_{g_{i,k}}\Phi_{i,n_i}(x_i)]^T.
	\end{aligned}
\end{equation*}
\textcolor{blue}{
The following assumption is employed to avoid the FET phenomenon. 
\begin{itemize}
	\item[{\bf A3:}] The internal dynamics $\eta_{i}$ in (\ref{mimo-linearized}) is ISS for all $i=1,\ldots,N$. 
\end{itemize}
}
Then the $r_{i,l}$-th equations, $l=1,\ldots,p$, in the linearized model (\ref{mimo-linearized}) of agents can be collected and written in the following form
\begin{equation}
	\label{collected-eq}
	\dot{\check{\xi}}_{i} = \check{\alpha}_{i}(\xi_{i},\eta_{i})+\Pi_{i}(\xi_{i},\eta_{i})u_i,
\end{equation}
where $\check{\xi}_{i}=[\xi_{i,r_{i,1}}^{1},\xi_{i,r_{i,2}}^{2},\ldots,\xi_{i,r_{i,p}}^{p}]^T \in \mathbb{R}^{p}$, 
$\check{\alpha}_{i}(\xi_{i},\eta_{i})=[\alpha_{i}^{1}(\xi_{i},\eta_{i}),\ldots,\alpha_{i}^{p}(\xi_{i},\eta_{i})]^T \in \mathbb{R}^{p}$, 
$\Pi_{i}(\xi_{i},\eta_{i}) \in \mathbb{R}^{p\times m_i}$ is defined in (\ref{decoupled-matrix}).  
Since ${\rm rank}(\Pi_{i}(\xi_{i},\eta_{i}))=p$, there exists a right inverse $\Pi_{i}(x_i)^{\dag}$ of $\Pi_{i}(x_i)$ defined by 
$\Pi_{i}(\xi_{i},\eta_{i})^{\dag} \triangleq \Pi_{i}(\xi_{i},\eta_{i})^T(\Pi_{i}(\xi_{i},\eta_{i})\Pi_{i}(\xi_{i},\eta_{i})^T)^{-1}.$  
Accordingly, the MIMO nonlinear system (agent) (\ref{mimo-nonlin-agent-aff}) can be input-output decoupled by the following controller
\begin{equation}
	\label{mimo-aff-decoupled-ctlr}
	u_i=\Pi_{i}(\xi_{i},\eta_{i})^{\dag}(-\check{\alpha}_{i}(\xi_{i},\eta_{i})+\check{u}_i),
\end{equation}
where $\check{u}_i=[\check{u}_{i,1},\ldots,\check{u}_{i,p}]^T \in \mathbb{R}^{p}$ is a new control input vector. 
As a result, the input-output decoupling is achieved for each agent as follows,
\begin{equation}
	\label{mimo-decoupled}
	\begin{aligned}
		\dot{\xi}_{i,1}^{l} &= \xi_{i,2}^{l}, \\
		\dot{\xi}_{i,2}^{l} &= \xi_{i,3}^{l}, \\
		 & \,~ \vdots  \\
		\dot{\xi}_{i,r_{i,l}}^{l} &= \check{u}_{i,l}, \\
		\dot{\eta}_{i} &= \vartheta_{i}(\xi_{i},\eta_{i})+\sum_{k=1}^{m_i}\chi_{i,k}(\xi_{i},\eta_{i})u_{i,k}, \\
		y_{i}^{l} &= \xi_{i,1}^{l}, \; \forall \; l=1,\ldots,p; i=1,\ldots,N.
	\end{aligned}
\end{equation}
At this point, we can see that the consensus design for output vectors of nonlinear agents (\ref{mimo-nonlin-agent-aff}) is decomposed into independent consensus designs of individual outputs of agents that is similar to SISO affine nonlinear MASs in Section \ref{siso-aff}. 
Hence, all steps of designing consensus controller for SISO affine nonlinear MASs can be adopted straightforwardly. 
Thus, to avoid the complexity and duplication in representing results, we skip the details here.
Similar situation applies to MIMO general nonlinear MASs. 
\section{\textcolor{blue}{Output Consensus of Heterogeneous Nonlinear MASs under Switching Topology} }
\label{consensus-mimo-nonlin-switch}

\textcolor{blue}{
In this section, we aim at investigating the consensus design for heterogeneous nonlinear MASs subjected to randomly switching topologies described by continuous-time Markov chains. 
Due to space limitation, only results for SISO affine nonlinear MASs are presented. }

\textcolor{blue}{
Since the communication topology is randomly switching, $\mathcal{G}$ is time-varying and is denoted by $\mathcal{G}(t)$. 
Suppose that $\mathcal{G}(t)$ switches among the elements of a finite set of $\ell$ topologies $\mathcal{S}_{\mathcal{G}} \triangleq \{\mathcal{G}_{1},\ldots,\mathcal{G}_{\ell}\}$, 
where the switching process is represented by a continuous-time Markov chain with a switching signal $\sigma(t) \in \{1,\ldots,\ell\}$. 
Denote $\mathcal{Q}=[q_{ij}] \in \mathbb{R}^{\ell \times \ell}$ and $\pi=[\pi_{1},\ldots,\pi_{\ell}]^T$ the transition rate matrix and the stationary distribution of this Markov process.  
Here, we assume that the Markov process is ergodic, so $\pi$ is unique, $\pi_{k}>0 \;\forall\; k=1,\ldots,\ell$,  and each state of the Markov chain can be reached from any other state. 
Furthermore, we can also assume that the Markov process starts from  $\pi$ \cite{You:2013}. 
As a result, the distribution of $\sigma(t)$ is equal to $\pi$ for all $t\geq 0$. 
} 

\textcolor{blue}{
Let us denote the union of all possible topologies by $\mathcal{G}_{\cup} \triangleq \displaystyle \bigcup_{k=1,\ldots,\ell}\mathcal{G}_{k}$. 
The following assumption is utilized. 
\begin{itemize}
	\item {\bf A4:} {\it $\mathcal{G}_{\cup}$ has a spanning tree and is balanced. }
\end{itemize}
The consensus of agents in this context is defined as follows. 
\begin{defi}
\label{ms-consensus}
The nonlinear MAS (\ref{siso-nonlin-agent}) with a randomly switching topology $\mathcal{G}(t)$ is said to reach a {\it mean-square consensus} for any initial condition of agents and any initial distribution of the continuous-time Markov process if
\begin{equation}
	\label{consensus-ms}
	\lim_{t \rightarrow \infty} \mathbb{E}\left[ \|x_{i}(t)-x_{j}(t)\|^{2} \right] = 0, \; \forall \; i,j=1,\ldots,N, 
\end{equation}
\end{defi}
where $\mathbb{E}[\cdot]$ denotes the expectation taken with some chosen probability measure. 
\begin{rem}
Note that assumption A4 is milder than the frequently used one in literature (see e.g., \cite{You:2013,Z.Li:2015}) which assumes the balance of $\mathcal{G}_{k} \;\forall\; k=1,\ldots,\ell$.  
This relaxation on the switching topologies is an advantage that greatly broaden the class of MAS topologies under random switches to achieve consensus. 
\end{rem}
}

\textcolor{blue}{
Consequently, the following theorem shows that our proposed distributed rank-$1$ consensus controller design in Section \ref{siso-aff} 
can be generalized to this scenario of switching topologies.
\begin{thm}
\label{switching-thm}
Under assumption A4, the nonlinear MAS (\ref{siso-nonlin-agent}) reaches a mean-square consensus in the sense of Definition \ref{ms-consensus} 
by the distributed dynamic consensus controller (\ref{consensus-ctlr-siso-aff}) with the following rank-$1$ cooperative input 
\begin{equation}
	\label{switching-ctlr-siso-aff-rank1}
	v_{i}(t) = -\mu \sqrt{q_1\hat{r}}\nu^T \sum_{j\in\mathcal{N}_i(t)}a_{ij}(t)(\hat{\xi}_{i}(t)-\hat{\xi}_{j}(t)). 
\end{equation} 
Moreover, the consensus speed is specified by
\begin{equation}
	\label{switching-speed}
	\pi^{\ast}\mu\sqrt{q_{1}\hat{r}}B^T\nu\lambda_{\min}\left(\mathcal{L}_{\cup}+\mathcal{L}_{\cup}^T\right),
\end{equation}
where $\pi^{\ast}=\displaystyle \min_{k=1,\ldots,\ell}\pi_{k}$, $\mathcal{L}_{\cup}$ is the Laplacian matrix associated with $\mathcal{G}_{\cup}$. 
\end{thm}
}

\begin{IEEEproof}
\textcolor{blue}{
Denote
\begin{align*}
\hat{\xi}_{\rm ave} &= \frac{1}{N}(\hat{\xi}_{1}(0)+\cdots+\hat{\xi}_{N}(0)), \\
\delta_{i}(t) &= \hat{\xi}_{i}(t)-\hat{\xi}_{\rm ave}, k=i,\ldots,N, \\
\delta(t) &= \left[\delta_{1}(t)^T,\ldots,\delta_{N}(t)^T\right]^T.
\end{align*}
It is then followed from substituting the  rank-$1$ cooperative input (\ref{switching-ctlr-siso-aff-rank1}) to the homogeneous linearized systems (\ref{subsys}) that 
$$\dot{\delta}(t)=[I_{N}\otimes A-\mu \mathcal{L}(t)\otimes(B\hat{r}B^TP_{1})]\delta(t).$$
Let us define the following Lyapunov functions
\begin{align*}
V(t) &= \mathbb{E}\left[ \delta(t)^T(I_{N}\otimes P_{1})\delta(t) \right], \\
V_{k}(t) &= \mathbb{E}\left[ \delta(t)^T(I_{N}\otimes P_{1})\delta(t)1_{\{\sigma(t)=k\}} \right], k=1,\ldots,\ell,
\end{align*}
where $1_{\cdot}$ is the Dirac measure. Obviously, $V(t)=\sum_{k=1}^{\ell}{V_{k}(t)}$. 
Consequently, using Lemma 4.2 in \cite{Fragoso:2005}, we obtain 
\begin{align*}
dV_{k}(t) &= \mathbb{E}\left[ (d\delta(t))^T(I_{N}\otimes P_{1})\delta(t)1_{\{\sigma(t)=k\}}  \right. \\
           & \left. \qquad \; +\delta(t)^T(I_{N}\otimes P_{1})d\delta(t)1_{\{\sigma(t)=k\}} \right]  \\
					 & \qquad \; +\sum_{j=1}^{\ell}{q_{jk}V_{j}(t)dt} + o(dt),
\end{align*}
where $o(\cdot)$ stands for the Little-o notation. Accordingly,
\begin{align*}
\dot{V}_{k}(t) &= \mathbb{E}\left[ \delta(t)^T[I_{N}\otimes (P_{1}A+AP_{1})  \right. \\
           & \left. \qquad \; -\mu(\mathcal{L}(t)+\mathcal{L}^T(t))\otimes(P_{1}B\hat{r}B^TP_{1})]\delta(t)1_{\{\sigma(t)=k\}} \right] \\
					 &= \mathbb{E}\left[ -\mu\delta(t)^T[(\mathcal{L}(t)+\mathcal{L}^T(t))\otimes(\nu q_{1} \nu^T)]\delta(t)1_{\{\sigma(t)=k\}} \right],	
\end{align*}
since $P_{1}A=\nu p_{1} \nu^TA=0$ as shown in the proof of Theorem \ref{siso-aff-coro}. 
Hence, 
\begin{align*}
\dot{V}(t) &= \mathbb{E}\left[ -\mu\delta(t)^T \left( \sum_{k=1}^{\ell}{\pi_{k}(\mathcal{L}_{k}+\mathcal{L}^T_{k})}\otimes(\nu q_{1} \nu^T) \right) \delta(t) \right] \\
           & \leq -\pi^{\ast}\mu \mathbb{E}\left[ \delta(t)^T [(\mathcal{L}_{\cup}+\mathcal{L}^T_{\cup})\otimes(\nu q_{1} \nu^T)] \delta(t) \right], 
\end{align*}
since $\pi_{k} \geq \pi^{\ast} \; \forall \; k=1,\ldots,\ell$, and $\sum_{k=1}^{\ell}{\mathcal{L}_{k}}=\mathcal{L}_{\cup}$. 
On the other hand, $\mathcal{L}_{\cup}+\mathcal{L}^T_{\cup}$ can be regarded as a Laplacian matrix of an connected undirected graph due to assumption A4. 
Therefore, there exists an orthogonal matrix $U\in\mathbb{R}^{N\times N}$ such that $\mathcal{L}_{\cup}+\mathcal{L}^T_{\cup}=U\Lambda U^T$ where $\Lambda=\mathrm{diag}\{\lambda_{i}\}_{i=1,\ldots,N}$ is a diagonal matrix whose diagonal elements are eigenvalues of $\mathcal{L}_{\cup}+\mathcal{L}^T_{\cup}$ and $U=\left[ \frac{1}{\sqrt{N}}\mathbf{1}_{N},U_{2} \right]$ with $U_{2}\in\mathbb{R}^{N\times(N-1)}$. 
Subsequently, 
\begin{align*}
	& \delta(t)^T [(\mathcal{L}_{\cup}+\mathcal{L}^T_{\cup})\otimes(\nu q_{1} \nu^T)] \delta(t)   \\
  &= \delta(t)^T(U\otimes I_{n})[\Lambda\otimes(\nu q_{1} \nu^T)](U^T\otimes I_{n})\delta(t) \\
	&= \delta(t)^T(U_{2}\otimes I_{n})\mathrm{diag}\{\lambda_{i}\nu q_{1} \nu^T\}_{i=2,\ldots,N}(U_{2}^T\otimes I_{n})\delta(t) \\
	&\geq \lambda_{\min}\left(\mathcal{L}_{\cup}+\mathcal{L}_{\cup}^T\right)\sqrt{q_{1}\hat{r}}B^T\nu\delta(t)^T(U_{2}\otimes I_{n})(I_{N-1}\otimes P_{1}) \\ 
	& \quad \times (U_{2}^T\otimes I_{n})\delta(t) \\
	&= \lambda_{\min}\left(\mathcal{L}_{\cup}+\mathcal{L}_{\cup}^T\right)\sqrt{q_{1}\hat{r}}B^T\nu\delta(t)^T(I_{N}\otimes P_{1})\delta(t).
\end{align*}
This leads to 
\begin{equation*}
	\dot{V}(t) \leq -\pi^{\ast}\mu\sqrt{q_{1}\hat{r}}B^T\nu\lambda_{\min}\left(\mathcal{L}_{\cup}+\mathcal{L}_{\cup}^T\right) V(t).
\end{equation*} 
Thus, $V(t)$ exponentially converges to $0$ with the speed $\pi^{\ast}\mu\sqrt{q_{1}\hat{r}}B^T\nu\lambda_{\min}\left(\mathcal{L}_{\cup}+\mathcal{L}_{\cup}^T\right)$, 
i.e., the mean-square consensus is achieved with the speed specified in (\ref{switching-speed}). 
} 
\end{IEEEproof}

\section{\textcolor{blue}{Numerical Example} }
\label{eg}

\textcolor{blue}{
To illustrate the proposed approach, let us consider a simple MAS composing of $5$ distinct SISO affine nonlinear agents described by the following dynamics,
\begin{itemize}
	\item {\bf Agent $1$:} $f_{1}(x_{1})=[-x_{1,1}-x_{1,1}^{5}+x_{1,2},x_{1,3},0]^T$, 
	$g_{1}(x_{1})=[0,0,1]^T$, $h_{1}(x_{1})=x_{1,2}$.	
	\item {\bf Agent $2$:} $f_{2}(x_{2})=[-x_{2,1}-x_{2,1}^{3}+x_{2,2},x_{2,3},0]^T$, 
	$g_{2}(x_{2})=[0,0,1]^T$, $h_{2}(x_{2})=x_{2,2}$.		
	\item {\bf Agent $3$:} $f_{3}(x_{3})=[x_{3,1}x_{3,2}+x_{3,1}x_{3,3},x_{3,2}^2+x_{3,3},x_{3,1}+x_{3,2}x_{3,3}]^T$, 
	$g_{3}(x_{3})=[1,0,0]^T$, $h_{3}(x_{3})=x_{3,2}$.	
	\item {\bf Agent $4$:} $f_{4}(x_{4})=[-4x_{4,1}-x_{4,1}^{3}+x_{4,2},x_{4,3},0]^T$, 
	$g_{4}(x_{4})=[0,0,1]^T$, $h_{4}(x_{4})=x_{4,2}$.		
	\item {\bf Agent $5$:} $f_{5}(x_{5})=[-2x_{5,1}-x_{5,1}^{5}+x_{5,2},x_{5,3},0]^T$, 
	$g_{5}(x_{5})=[0,0,1]^T$, $h_{5}(x_{5})=x_{5,2}$.		
\end{itemize}
It can be verified that the relative degrees of agents $1,2,4,5$ are $2$ and of agent $3$ is $3$, then the maximum relative degree of agents is $3$ and hence by following the consensus design  
in Section \ref{siso-aff}, agents $1,2,4,5$ will be equipped with dynamic consensus controllers whereas agent $3$ will be incorporated with a static consensus controller. }

\textcolor{blue}{
{\bf For agent $i, i \in \{1,2,4,5\}$:} $\alpha_{i}(\xi_{i},\eta_{i})=0$, $\beta_{i}(\xi_{i},\eta_{i})=1$. 
The matrices of local dynamic controller is: $D_{i}=\begin{bmatrix} 0 & -b_{2} \end{bmatrix}$; $E_{i}=-b_{3}$; $G_{i}=1$; $H_{i}=1/\beta_{i}(\xi_{i},\eta_{i})$. }

\textcolor{blue}{
{\bf For agent $3$:} $\alpha_{3}(\xi_{3},\eta_{3})=x_{3,1}(x_{3,2}+x_{3,3})+(6x_{3,2}^{2}+3x_{3,3})(x_{3,2}^2+x_{3,3})+3x_{3,2}(x_{3,1}+x_{3,2}x_{3,3})$, $\beta_{3}(\xi_{3},\eta_{3})=1$. 
}

\subsection{ \textcolor{blue}{Fixed Directed Topology} }

\begin{figure}[thpb]
      \centering
      \includegraphics[scale=0.35]{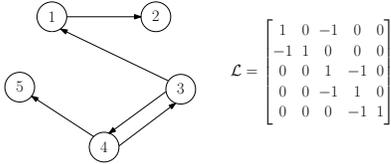}
      \caption{\textcolor{blue}{Communication structure in the MAS.} }
      \label{5agents-directed}
   \end{figure}

\textcolor{blue}{
The communication topology among agents and the associated Laplacian matrix in this case are presented in Figure \ref{5agents-directed}. }

\textcolor{blue}{	
Consequently, we would like to demonstrate the distributed minimum rank, i.e., rank-$1$ consensus controller in Theorem \ref{siso-aff-coro} that shows the advanced features of our proposed approach. 
Let us choose $b_{2}=2$ and $b_{3}=3$ then matrix $A$ of homogeneous linearized system has only one zero eigenvalue with the associated left eigenvector 
$\nu^T=\begin{bmatrix} 2 & 3 & 1 \end{bmatrix}$. 
}

\begin{figure}[thpb]
      \centering
      \includegraphics[scale=0.45]{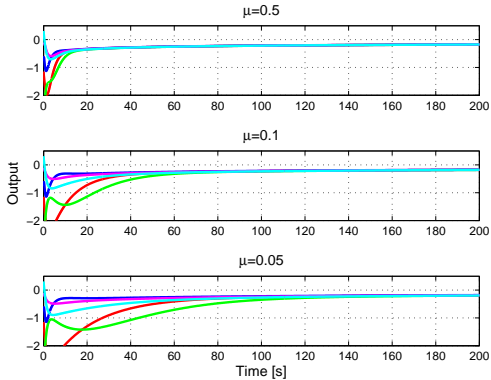}
      \caption{\textcolor{blue}{Consensus of the given nonlinear MAS by a distributed rank-$1$ controller with non-conservative coupling strength $\mu$.}}
      \label{mu_change}
   \end{figure}

\textcolor{blue}{
First, we attempt to verify the coupling strength $\mu$ to the consensus of agents by selecting $\hat{r}=1$, $q_{1}=1$ and varying $\mu$. 
The simulation results for the rank-$1$ consensus controller in this case are displayed in Figure \ref{mu_change}. 
It can be seen that even when $\mu$ is very small, the consensus among agents is still achieved. 
This confirms our claim on the non-conservative design of arbitrarily small but positive coupling strength. 
Furthermore, the consensus speed is slower as $\mu$ is smaller. That is because it solely depends on $\mu$ when $\mu$ is varied and small, 
which is deduced from the expression of consensus speed (\ref{consensus-speed-aff}). 
} 

\begin{figure}[thpb]
      \centering
      \includegraphics[scale=0.45]{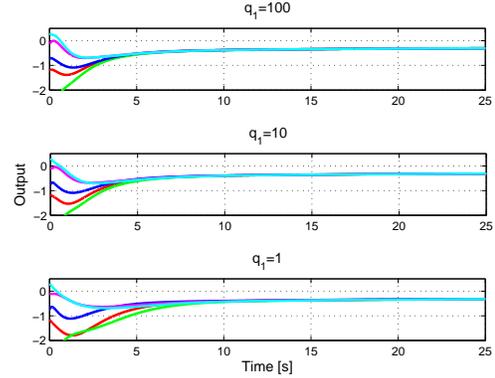}
      \caption{\textcolor{blue}{Consensus of the given nonlinear MAS by a distributed rank-$1$ controller as $q_{1}$ is changed.} }
      \label{q1_change}
   \end{figure}

\textcolor{blue}{
Next, we would like to check the effects of the parameters $q$ and $\hat{r}$ to the consensus speed. 
Since the roles of $q$ and $\hat{r}$ are similar in the consensus speed formula (\ref{consensus-speed-aff}), let us choose $\hat{r}=1$, $\mu=1$, and change $q_{1}$. 
Then we can observe from simulation results in Figure \ref{q1_change} that the consensus speeds of agents as $q_{1}=10$ and $q_{1}=10$ are similar and are faster than when $q_{1}=1$. 
This is explained by the consensus speed determined in (\ref{consensus-speed-aff}) as follows. 
We have $\lambda_{\min}(-A)=2$ while $\mu\sqrt{q_{1}\hat{r}}B^T\nu\mathrm{Re}(\lambda_{\min}(\mathcal{L}))=1.5 \sqrt{q_1}$, and hence when $q_{1}=1$ the consensus speed is equal to $1.5\sqrt{q_1}=1.5$, 
but when $q_{1}=10$ or $q_{1}=100$ the consensus speed is equal to $\lambda_{\min}(-A)=2$ which is independent of $q_{1}$. 
}

\subsection{ \textcolor{blue}{Switching Directed Topology} }

\begin{figure}[thpb]
      \centering
      \includegraphics[scale=0.35]{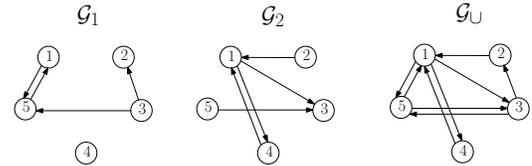}
      \caption{\textcolor{blue}{Switching topologies of the given nonlinear MAS.} }
      \label{5agents-switching}
   \end{figure}

\begin{figure}[thpb]
      \centering
      \includegraphics[scale=0.45]{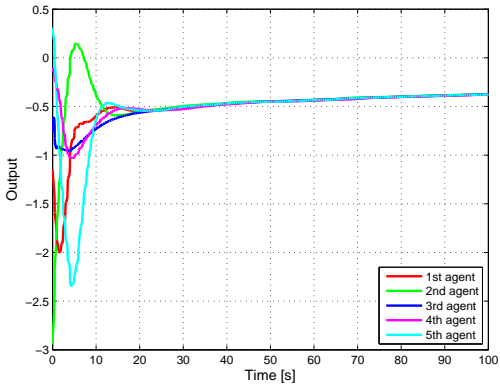}
      \caption{\textcolor{blue}{Consensus of the given nonlinear MAS under a randomly switching topology by a distributed rank-$1$ controller.} }
      \label{switching}
   \end{figure}

\textcolor{blue}{
Here we assume that the communication topology among agents is randomly switched between two directed graphs $\mathcal{G}_{1}$ and $\mathcal{G}_{2}$ shown in Figure \ref{5agents-switching}, 
where the random process is described by a continuous-time Markov chain with generator matrix 
$Q=\begin{bmatrix} -1 & 1 \\ 1 & -1 \end{bmatrix}$ 
and the invariant distribution $\pi=[\frac{1}{2}, \frac{1}{2}]$. 
It can be seen that neither $\mathcal{G}_{1}$ nor $\mathcal{G}_{2}$ is balanced and none of them has a spanning tree, but their union graph $\mathcal{G}_{\cup}$ is balanced and has a spanning tree. 
Next, the parameters of the distributed rank-$1$ consensus controllers are $\hat{r}=1$, $\mu=1$, $q=1$. 
Then the simulation result is exhibited in Figure \ref{switching}. We can observe that the outputs of nonlinear agents still reach a consensus in spite of the switching topology. 
This confirms our result in Theorem \ref{switching-thm}. 
}

\section{\textcolor{blue}{Conclusions and Discussions} }
\label{sum}

\textcolor{blue}{
This article has proposed a systematic framework to design distributed dynamic rank-$1$ consensus controllers for a fairly general class of heterogeneous MIMO nonlinear MASs subjected to fixed and randomly switching directed topologies.  
The framework has been developed based on the input-output feedback linearization and LQR methods with the following appealing properties. 
First, distributed {\it dynamic} consensus controllers are derived for {\it heterogeneous MIMO nonlinear MASs with arbitrary vector relative degree}. 
Second, the coupling strength in the consensus controller can be {\it arbitrarily small but positive} which allows us to achieve consensus with any speed. 
Third, the dynamic consensus controller has {\it minimum rank}, i.e., rank-$1$  which is very computationally efficient. 
And last, the proposed design works well under randomly switching topologies where the switched graphs are {\it unnecessary to be balanced}, which greatly relaxes the assumptions on switching topologies. 
}

The current results can be further developed in several directions that are worth investigating. One issue is the robustness and adaptability of the consensus controller in the presence of time delays, unmeasured disturbances or noises, and model uncertainties. Another direction is to design dynamic consensus controllers for heterogeneous nonlinear MASs under some constraints for control inputs or state flows.

\section*{Acknowledgment}

The author would like to sincerely thank the anonymous reviewers for their valuable comments and suggestions that significantly help to improving the quality of the paper. 

\ifCLASSOPTIONcaptionsoff
  \newpage
\fi



%

\bibliographystyle{IEEEtran}

\bibliography{References}

%
%

%

%
%
%
%
%
%
%
%
%
\end{document}